\documentclass[useAMS,senatbib]{mn2e}
\usepackage{graphics,graphicx}
\usepackage{psfig}
\def\ltapprox{\raise 2pt \hbox {$<$} \kern-1.1em \lower 5pt \hbox {$\approx$}}

\def\ltsim{\; \raise0.3ex\hbox{$<$\kern-0.75em \raise-1.1ex\hbox{$\sim$}}\; }
\def\gtsim{\; \raise0.3ex\hbox{$>$\kern-0.75em \raise-1.1ex\hbox{$\sim$}}\; }

\def\ie{{\it i.e.,~}}

\def\eg{{\it e.g.,~}}

\title[New scaling relations in cluster radio halos and the re-acceleration model]{New scaling relations in cluster radio halos and the re-acceleration model}

\author[R. Cassano et al.]{R. Cassano,$^{1,2}$\thanks{E-mail:
rcassano@ira.inaf.it} G. Brunetti$^{2}$, G. Setti$^{1,2}$, F Govoni$^{3}$, K. Dolag$^{4}$\\
$^1$ Dipartimento di Astronomia,Universita' di Bologna, via Ranzani 1, I-40127 Bologna, Italy\\
$^2$ INAF - Istituto di Radioastronomia, via P. Gobetti 101,I-40129 Bologna, Italy\\
$^3$ INAF - Osservatorio Astronomico di Cagliari, Loc. Poggio dei Pini, Strada 54, 09012 Capoterra, Italy\\
$^4$ Max-Planck Institut fur Astrophysik, Karl-Schwarzschild Strasse 1, D-85748 Garching, Germany}

\begin{document}


\pagerange{\pageref{firstpage}--\pageref{lastpage}} \pubyear{2006}

\maketitle

\label{firstpage}

\begin{abstract}

In this paper we derive new expected scaling relations for 
clusters with giant radio halos in the framework of the re-acceleration
scenario in a simplified, but physically motivated, form, namely: 
radio power ($P_{R}$) vs
size of the radio emitting region ($R_H$), and 
$P_{R}$ vs total cluster mass ($M_H$) contained in the emitting region 
and cluster velocity dispersion ($\sigma_H$) in this region.

We search for these correlations by analyzing the 
most recent radio and X-ray data available in the literature for a well known 
sample of clusters with giant radio halos. 
In particular we find a good correlation between $P_{R}$ and $R_H$ and a very
tight ``geometrical'' scaling between $M_H$ and $R_H$. From these correlations $P_R$ is also expected to scale with $M_H$ and $\sigma_H$ and this is confirmed by our analysis.
We show that all the observed trends can be
well reconciled with expectations in the case of a slight variation of the mean magnetic 
field strength in the radio halo volume with $M_H$.
A byproduct correlation between $R_H$ and $\sigma_H$ is also found, and can be further
tested by optical studies.
In addition, we find that observationally $R_H$ scales non-linearly
with the virial radius of the host cluster, and this immediately means that the fraction 
of the cluster volume which is radio emitting increases with cluster mass 
and thus that the non-thermal component in clusters is not self-similar. 
\end{abstract}


\section{Introduction}

Radio halos are diffuse Mpc scales synchrotron radio sources
observed at the center of a growing number ($\sim$ 20) of massive
galaxy clusters (see \eg Feretti 2005 for a review).
Radio halos are always found in merging clusters (\eg Buote 2001; Schuecker et
al 2001) thus suggesting a link between the dynamical status of clusters
and the radio halos.
Observations show that radio halos are rare; however present data suggest that their detection
rate increases with increasing the X-ray luminosity
of the host clusters and reaches 30-35\% for galaxy clusters at $z\le 0.2$ and
with X-ray luminosity larger than $10^{45}$ $h_{50}^{-1}$ erg/s (Giovannini, Tordi \& Feretti
1999, GTF99).

Two main possibilities have been so far investigated to explain the radio halos:
{\it i)} the so-called {\it re-acceleration} models, whereby
relativistic electrons injected in the intra cluster medium (ICM) are
re-energized {\it in situ} by various mechanisms associated with the
turbulence generated by massive merger events (\eg Brunetti et al. 2001;
Petrosian et al. 2001); {\it ii)} the {\it secondary electron} models, whereby the
relativistic electrons are
secondary
products of the hadronic interactions of
cosmic rays with the ICM (\eg Dennison 1980; Blasi \& Colafrancesco 1999).

\noindent Recently, calculations in the framework of the {\it re-acceleration} scenario
have modelled the connection between radio halos and cosmological cluster mergers,
and investigated the observed correlations between the synchrotron radio
power and the X-ray properties of the hosting clusters (Cassano \& Brunetti 2005, CB05;
Cassano, Brunetti \& Setti 2006, CBS06). Observed correlations relate the radio power at
1.4 GHz ($P_{1.4}$) with the X-ray luminosity ($L_X$), temperature ($T$) and cluster mass
(Liang 1999; Colafrancesco 1999; Feretti 2000,2003; Govoni et al. 2001a; En\ss lin and R\"ottgering 2002; CBS06); also a trend between the largest linear size of radio halos and the X-ray luminosities of the hosting clusters is found (Feretti 2000).
In particular, CBS06 found a correlation between $P_{1.4}$ and the virial
mass $M_v$ of the hosting clusters, $P_{1.4}\propto M_v^{2.9\pm0.4}$, by combining
the $P_{1.4}-L_X$ correlation derived from a sample of 17 giant radio halos with
the $M_v-L_X$ correlation obtained for a large sample of galaxy cluster compiled by
Reiprich \& B\"oheringer (2002). However, this correlation, which has been discussed in the
particle re-acceleration scenario by CBS06, relates quantities which pertain to very different spatial regions: the observed radio emission comes from a radial size $R_H\sim3-6$ time smaller than the virial radius $R_v$.

\noindent In this paper we discuss expected scaling relations for radio halos in the framework of
the re-acceleration scenario {\it in its simplest form}.
Then, we derive a novel observed correlation between the radio power of radio halos and their extension and a tight ``geometrical'' correlation between the size of radio halos and the mass of the cluster within the emitting region. We also present additional correlations 
which are expected on the basis of these two scalings.
Finally we compare all these observed correlations with the model expectations.

A $\Lambda$CDM ($H_{o}=70$ km $s^{-1}$ Mpc$^{-1}$, $\Omega_{m}=0.3$, $\Omega_{\Lambda}=0.7$) cosmology is adopted.

\section{Particle acceleration scenario}

\subsection{Main features and implications of the re-acceleration model}

The particle re--acceleration model is designed to explain the origin
of the synchrotron radio emission diffused on scales larger than that
of the cluster cores (giant radio halos), while the so called mini--halos
and other smaller scale diffuse sources at the cluster center (e.g. core 
halo sources) might have a different origin (\eg Gitti, Brunetti, Setti 2002; Pfrommer
\& Ensslin 2004, and ref. therein).

In the conventional particle re--acceleration scenario the lower energy
electrons ($\gamma\sim 100-300$), relic of past 
activities in the clusters, are re-energized due to resonant 
and/or non-resonant interactions with the turbulence 
developed during cluster-cluster mergers.
Turbulence and shear flows are expected to amplify the magnetic 
field in galaxy clusters (\eg Dolag et al. 2002, 2005; Br\"uggen et al. 2005) however, the decay 
time-scale of the magnetic field is expected to be larger than several Gyr (\eg Subramanian, Shukurov and Haugen 2006) and thus the particle re-acceleration process can be thought as 
occurring in a stationary magnetic field amplified during the previous merging history of the cluster.

The basic features of this model can be briefly summarized 
as follows: 

a) The average synchrotron spectrum of radio halos is curved and can be 
approximated by a relatively steep quasi--power law which 
further steepens at higher frequencies
up to a cut-off  frequency.

The curved, cut-off spectrum is a unique feature of the 
re-acceleration model, which well represents
the typical observed radio halo spectrum, due to the existence of 
a maximum energy of the radiating
electrons (at $\gamma_{max} < 10^5$) determined by the
balance between the energy gains (re--acceleration processes) and synchrotron 
and inverse Compton losses (\eg Brunetti et al. 2001, 2004; Ohno et al 2002; 
Kuo et al. 2003). Accordingly, the detection of a radio halo critically
depends on cut--off frequency which 
should be sufficiently larger than the observing frequency.
As a consequence, there is a threshold in the efficiency which should
be overcome by the re--acceleration processes in order to accelerate the electrons at the 
energies necessary to produce radio emission at the observed frequency in the clusters' magnetic fields. 
In the merger--related scenario it is expected that only mergers between 
massive galaxy clusters may be able to generate enough turbulence on large 
scales to power giant radio halos at GHz frequencies, thus 
not all clusters which show some merger activity are expected to possess a
giant radio halo. 
In particular, CB05 show that the expected fraction of clusters with radio halos 
increases with cluster mass due to a more efficient particle re-acceleration 
process in more massive galaxy clusters, and this is 
in line with the increase of the fraction of radio halos with cluster mass
which is claimed from the analysis of present radio surveys (\eg GTF99).

b) In the re--acceleration model radio halos should be transient phenomena 
in dynamically disturbed clusters. The time scale of the radio halo phenomena comes 
from the combination of the time necessary for the cascading of the turbulence 
from cluster scales to the smaller scales relevant for particle acceleration, 
of the time--scale for dissipation of the turbulence and of the cluster--cluster 
crossing time.

Present observations suggest that radio halos are preferentially found in 
dynamically disturbed systems (\eg Buote 2001; Govoni et al. 2004).
Under the hypothesis that radio halos form in merging clusters in the 
hierarchical scenario, Kuo et al. (2004) found that the lifetime of these 
radio halos should be $\ltsim 1$ Gyr to not overproduce the 
observed occurrence of these sources.

\subsection{Predicted scalings for giant radio halos}

In this Section we derive scaling expectations for giant and powerful
radio halos in the context of the re--acceleration scenario in its simplest form.

The most important ingredient is the energy of the turbulence
injected in the ICM. Numerical simulations of merging clusters show that 
infalling sub-halos induce turbulence (\eg Roettiger, Loken \& Barns 1997;
Ricker \& Sarazin 2001; Tormen, Moscardini \& Yoshida 2004). 
An estimate of the energy of merging-injected turbulence has been recently 
derived in CB05 by assuming that a fraction of the PdV work done by the infalling sub-halos is injected into compressible turbulence. 
They show that the turbulent energy is expected to roughly scale
with the thermal energy of the ICM, a result in line with recent
analysis of numerical simulations (Vazza et al. 2006, V06).

Once injected this turbulence is damped by Transit-Time-Damping (TTD)
resonance with thermal and relativistic particles (at a rate
$\Gamma_{th}$ and $\Gamma_{rel}$, respectively). 
Since the damping time is shorter than the other relevant time scales
(dynamical and re-acceleration) the energy 
density of the turbulence reaches a stationary condition given by 
$\dot{\varepsilon}_{t}/(\Gamma_{th}+\Gamma_{rel})$, where
$\dot{\varepsilon}_{t}$ is the turbulence injection rate (CB05). 
When re--acceleration starts, the bulk of the energy density
of compressible modes which is damped by the relativistic
electrons goes into the re--energization of these electrons.
On the other hand, after a few re--acceleration times, in a time--scale
of the order of the typical age of radio halos, electrons are boosted at high
energies at which radiative losses are severe ($\propto E^2$) and the effect 
of particle re--acceleration ($\propto E$) is balanced by that of 
radiative losses. 
The electron spectrum gradually approaches a quasi-stationary condition and it can be assumed that
the energy flux of the turbulent modes {which goes into relativistic electrons is essentially re--radiated via synchrotron and inverse Compton mechanisms:

\begin{equation}  
(\frac{\dot{\varepsilon}_{t}\,\Gamma_{rel}}{\Gamma_{th}+\Gamma_{rel}}) 
\propto (\dot{\varepsilon}_{syn}+\dot{\varepsilon}_{ic})\qquad \Rightarrow
\dot{\varepsilon}_{syn} \propto \frac{\dot{\varepsilon}_t 
\times (\Gamma_{rel}/{\Gamma_{th})}} 
{(1+\frac{\dot{\varepsilon}_{ic}}{\dot{\varepsilon}_{syn}})}
\label{epsilont}
\end{equation}
 
\noindent where $\dot{\varepsilon}_{syn}$ and $\dot{\varepsilon}_{ic}$ are the 
synchrotron and IC emissivities (and $\Gamma_{th}>>\Gamma_{rel}$, CB05; 
Brunetti \& Lazarian 2007).

\noindent 
The ratio $\dot{\varepsilon}_{ic}/\dot{\varepsilon}_{syn}$ simply depends on
$(B_{cmb}/B_{H})^2$, where $B_{cmb}=3.2\,(1+z)^2\,\mu$G is the equivalent
magnetic field strength of the CMB (z, the redshift) and $B_{H}$ the 
mean magnetic field strength in the radio halo volume, which can 
be parameterized as $B_{H}\propto M_H^{b_H}$
with $M_H$ the total cluster mass within $R_H$ (the average radius of the radio emitting region). 

\begin{figure*}
\includegraphics[width=0.4\textwidth]{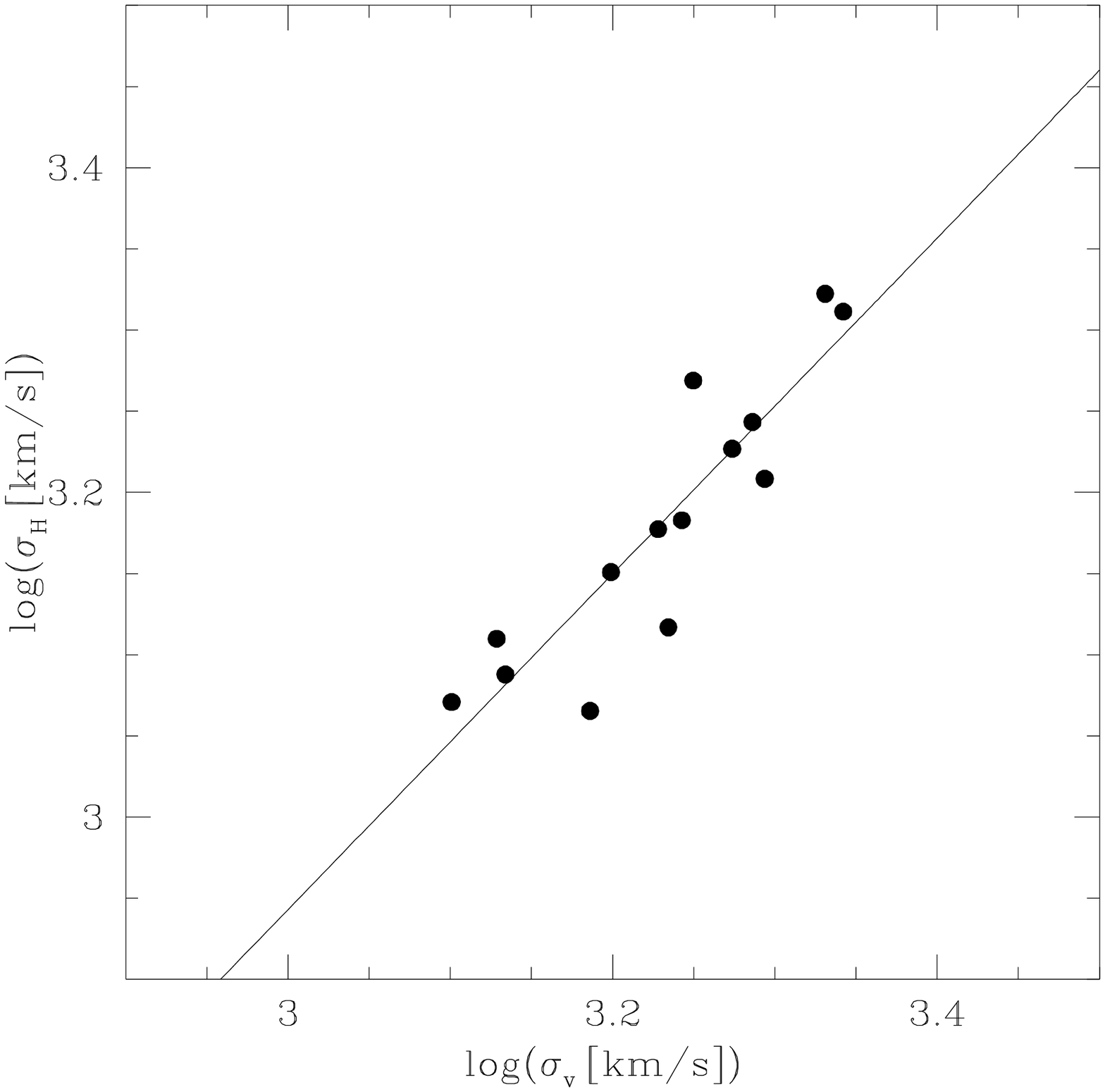}
\includegraphics[width=0.4\textwidth]{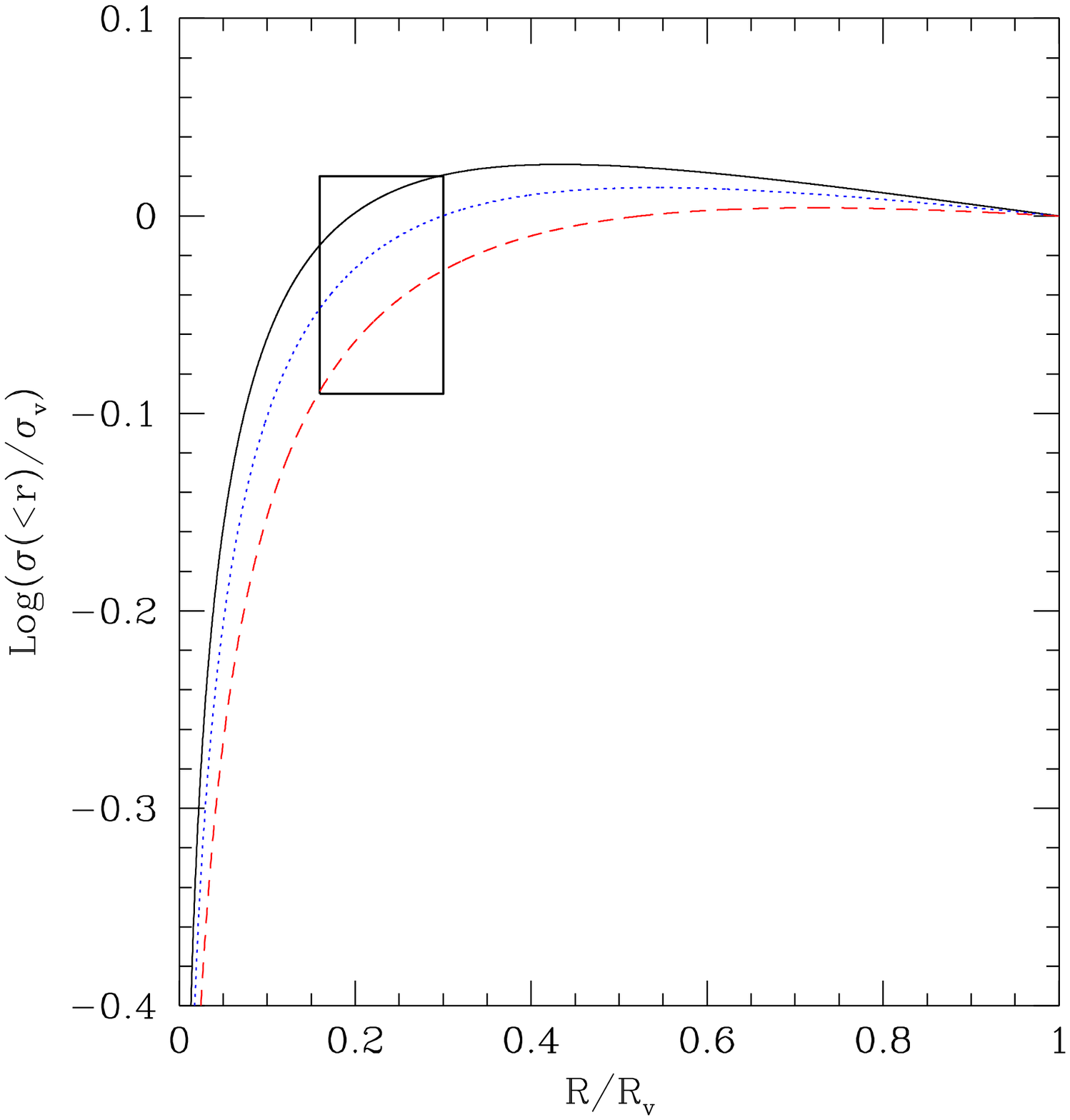}
\caption[]{ a) $\sigma_H$ versus virial velocity dispersion for the galaxy clusters
in our sample together with the best fit power-law $\sigma_H\propto\sigma_v^{1.03}$; 
b) Circular velocity profile, $\sigma(<r)=(GM(<r)/r)^{0.5}$, normalized to the virial value
from Navarro-Frenk-White (NFW; Navarro, Frenk and White 1997) models with $c=R_v/r_s=3,4,5$ for the dashed, dotted and solid lines, respectively.
The rectangle indicates the region of the radio halos: $R_H/R_v\sim 0.16-0.3$. $\sigma_H/\sigma_v$ varies by a factor of less than 15\% (for a fixed $c$) in our sample.}
\label{sigmaH_sigmav}
\end{figure*}

Based on CB05, the injection rate of the turbulence in the radio halo volume 
can be estimated as $\dot{\varepsilon}_t \propto \overline{\rho}_{H}\,v_i^2/\tau_{cros}$,
where $\overline{\rho}_{H}$ is the mean density of the ICM in the radio halo volume, $v_i$ is the 
cluster-cluster impact velocity, $v_i^2\propto M_v/R_v$, and $\tau_{cros}\propto (R_v^3/M_v)^{0.5}$ 
is the cluster-cluster crossing time (see CB05) and is constant by 
definition of virial mass in the cosmological hierarchical model (\eg Borgani 2006, for a review). 
In the case $R_H$ is larger than the cluster core radius it is $v_i^2\propto M_v/R_v \propto M_H/R_H$ 
and $\sigma_H$, the velocity dispersion inside $R_H$, is $\sigma_H\equiv G\,M_H/R_H\approx \sigma_v^2$ 
(for the sake of clarity in Fig.~\ref{sigmaH_sigmav} we report a comparison between $\sigma_H$ and $\sigma_v$ 
for our sample of clusters with radio halos). Thus we shall simply assume that the injection rate of turbulence in the 
radio halo volume is $\dot{\varepsilon}_t \propto \overline{\rho}_{H}\,\sigma_H^2$. The term $\Gamma_{rel}/\Gamma_{th}$ is
$\propto\epsilon_{rel}/\epsilon_{th} \times \sqrt[]{T}$ (Brunetti 2006, 
Brunetti \& Lazarian 2007), where $T$ is the temperature of the cluster gas, 
and $\epsilon_{rel}/\epsilon_{th}$ is the ratio between the energy densities in 
relativistic particles and in the thermal plasma. Although this ratio might reasonably
vary from cluster to cluster, we shall assume that it does not 
appreciably change in any {\it systematic} way with cluster mass 
(or temperature), at least if one restricts to the relatively
narrow range in cluster mass spanned by clusters with giant radio halos
(see also the results from numerical simulations for cosmic rays in Jubelgas et al. 2006).
Then from Eq.\ref{epsilont} the total emitted radio 
power is: 

\begin{eqnarray}
P_R=\int\dot{\varepsilon}_{syn}\,dV_H \propto \frac{M_H\,\sigma_H^3}
{\mathcal{F}(z,M_H,b_H)}
\label{epsilonsynint}
\end{eqnarray}

\noindent 
where we have taken $\sqrt[]{T}\propto \sigma_H$ and 
$\mathcal{F}(z,M_H,b_H)=\big[1+(3.2\,(1+z)^2/B_H)^2\big]$.
The expression $\mathcal{F}$ (Fig.\ref{F_MH}) is constant in the asymptotic 
limit $B_H^2>>B_{cmb}^2$ or in the simple case in which the rms magnetic
field in the radio halo region is independent of the cluster mass.
For $B_H^2 << B_{cmb}^2$ one has that $\mathcal{F}^{-1} \propto
M_H^{2 b_H}$, thus in the general case the expected scaling
is steeper (slightly for $B_H$ of the order of 
a few $\mu$G) than that obtained by assuming a constant $\mathcal{F}$.

\begin{figure}
\centerline{\psfig{figure=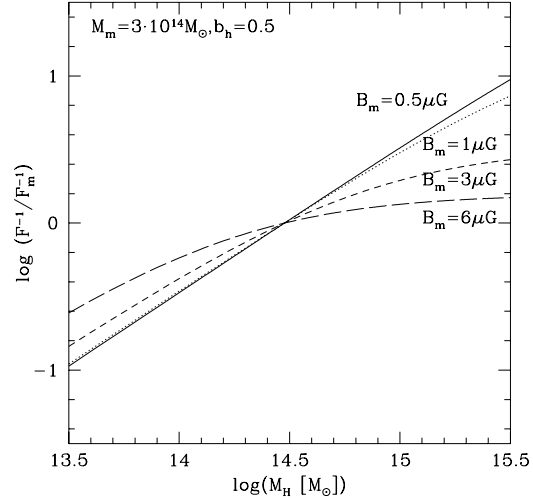,width=7cm}}
\caption[]{Function $\mathcal{F}^{-1}$, normalized to the $\mathcal{F}^{-1}$ value 
for a mean $M_H=M_m=3\cdot 10^{14}\,M_{\odot}$, as a function of $M_H$, for
$b_H=0.5$ and assuming different values of the magnetic field $B_H$ corresponding to the mean mass $B_m=0.5,\, 1,\, 3,\, 6\,\mu$G, from top to bottom.}
\label{F_MH} 
\end{figure}

It is important to stress here that the expression in Eq.\ref{epsilonsynint} is a general
theoretical trend which implies simple scaling relations. Indeed, by taking $\sigma_H \approx \sqrt{G M_H / R_H}$ and under the assumption that the mass scales with $R_H$
as $M_H\propto R_H^{\alpha}$ (see also Sect.~3.2), Eq.\ref{epsilonsynint} (with $\mathcal{F}\sim$ constant) 
entails the correlations:
 
{\setlength\arraycolsep{2pt} 
\begin{eqnarray}
\label{eq.RH}
P_R&\propto &R_H^{\frac{5\alpha-3}{2}} \\
\label{eq.MH}
P_R&\propto &M_H^{\frac{5\alpha-3}{2\alpha}} \\
\label{eq.sigmaH}
P_R&\propto &\sigma_H^{\frac{5\alpha-3}{\alpha-1}}
\end{eqnarray}} 

\noindent the effect of a non constant $\mathcal{F}$ is a steepening
(although not substantial for $\sim \mu$G fields) 
of these scalings.

\section{Observed scaling relations in clusters with radio halo}

Motivated by the theoretical expectations outlined 
in the previous Section, 
we have searched for the predicted scaling relations 
in the available data set for giant radio halos.
Operatively, we will first discuss the case of the $P_R-R_H$ scaling expected
in Eq.\ref{eq.RH}, which will allow us to address the tricky point of the measure of $R_H$ in radio halos,
and then we will show that a tight observational $R_H-M_H$ scaling exists for radio halos. Then, we will discuss and verify the byproduct observational scalings between $P_R-M_H$ and $P_R-\sigma_H$.

We consider a sample of 15 clusters with known giant radio halos ($R_H\gtsim 300$ kpc) 
already analyzed in CBS06, with the exclusion of CL0016+16, due to the lack of good radio images to measure $R_H$, and of A754, due to very complex radio structure.  
References for 14 giant radio halos are given in CBS06, while for A2256 we use the more recent radio data from Clarke \& En\ss lin (2006). In Tab.\ref{tab.RH} we report the relevant observed and derived quantities for our sample. 

\begin{figure}
\centerline{\psfig{figure=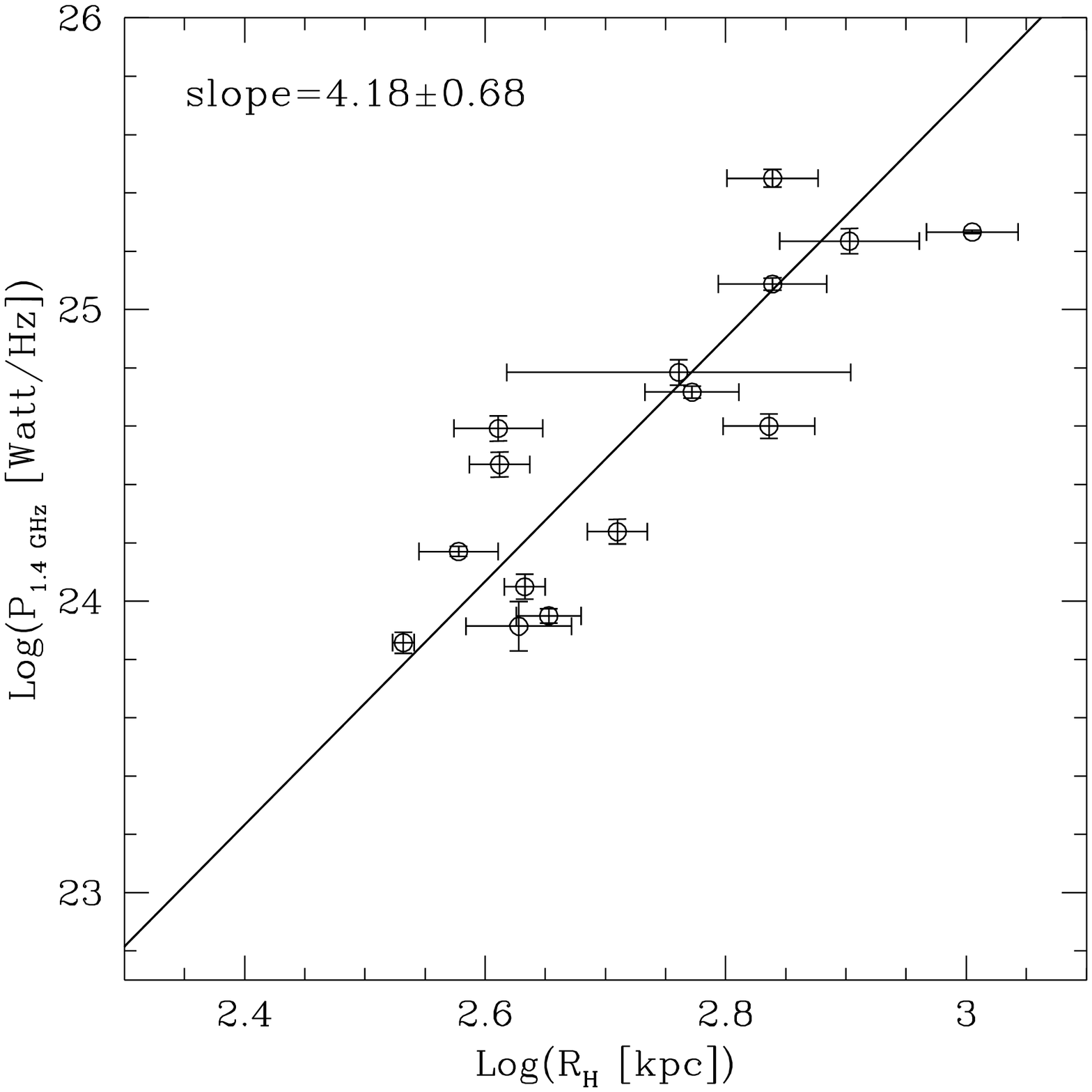,width=7cm}}
\caption[]{$P_{1.4}$ versus $R_H$. The fit has been performed using a power-law form 
in the log-log space and the best fit slope is reported in the panel.}
\label{Lr_RH}
\end{figure} 

\subsection{Radio power versus sizes of radio halos}

\begin{table*}
\begin{center}
\caption{In 
Col.(1): Cluster name. 
Col.(2): cluster redshift. 
Col.(3): logarithm of the radio power at $1.4$ GHz, $P_{1.4}$, in unit of Watt/Hz.
Col.(4): logarithm of the size of the radio halos, $R_H$, in unit of kpc $h_{70}^{-1}$.  
Col.(5): logarithm of the total cluster mass inside $R_H$, $M_H$, in unit of solar masses.
The references for the cluster redshift and radio power are reported in CBS06, while for A2256 we use the more recent radio data from Clarke \& En\ss lin (2006). }
\begin{tabular}{lccccc}
\hline
\hline
cluster's  & z  & Log($P_{1.4}$)&      Log($R_H$)     & Log($M_H$)   & Log($\sigma_H^2$) \\
  name     &    & [Watt/Hz]     & [kpc $h_{70}^{-1}$] & [$M_{\odot}\,h_{70}^{-1}$]& [$km^2\, s^{-2}$] \\
\hline
\hline
1E50657-558  &  0.2994 & $ 25.45\pm  0.03 $  & $2.84\pm 0.04$ &  $  14.83\pm  0.07 $ & $ 6.63\pm  0.08$  \\  
A2163        &  0.2030 & $ 25.27\pm  0.01 $  & $3.01\pm 0.04$ &  $  15.02\pm  0.05 $ & $ 6.65\pm  0.07$ \\  
A2744        &  0.3080 & $ 25.23\pm  0.04 $  & $2.90\pm 0.06$ &  $  14.76\pm  0.10 $ & $ 6.49\pm  0.11$ \\  
A2219        &  0.2280 & $ 25.09\pm  0.02 $  & $2.84\pm 0.05$ &  $  14.66\pm  0.08 $ & $ 6.46\pm  0.09$ \\  
A1914        &  0.1712 & $ 24.72\pm  0.02 $  & $2.77\pm 0.04$ &  $  14.68\pm  0.05 $ & $ 6.54\pm  0.06$ \\  
A665         &  0.1816 & $ 24.60\pm  0.04 $  & $2.84\pm 0.04$ &  $  14.57\pm  0.09 $ & $ 6.37\pm  0.10$ \\  
A520         &  0.2010 & $ 24.59\pm  0.04 $  & $2.61\pm 0.04$ &  $  14.21\pm  0.10 $ & $ 6.24\pm  0.11$ \\ 		 
A2254        &  0.1780 & $ 24.47\pm  0.04 $  & $2.61\pm 0.03$ &        $--$                &     $--$                 \\
A2256        &  0.0581 & $ 23.91\pm  0.08 $  & $2.63\pm 0.04$ &  $  14.17\pm  0.09 $ & $ 6.18\pm  0.11$   \\   
A773         &  0.2170 & $ 24.24\pm  0.04 $  & $2.71\pm 0.03$ &  $  14.43\pm  0.05 $ & $ 6.36\pm  0.06$   \\  
A545         &  0.1530 & $ 24.17\pm  0.02 $  & $2.58\pm 0.03$ &  $  14.08\pm  0.30 $ & $ 6.13\pm  0.30$   \\  
A2319        &  0.0559 & $ 24.05\pm  0.04 $  & $2.63\pm 0.02$ &  $  14.30\pm  0.03 $ & $ 6.30\pm  0.03$   \\  
A1300        &  0.3071 & $ 24.78\pm  0.04 $  & $2.76\pm 0.14$ &  $  14.54\pm  0.17 $ & $ 6.42\pm  0.22$   \\  
Coma (A1656)        &  0.0231 & $ 23.86\pm  0.04 $  & $2.53\pm 0.01$ &  $  14.12\pm  0.03 $ & $ 6.22\pm  0.03$   \\  
A2255        &  0.0808 & $ 23.95\pm  0.02 $  & $2.65\pm 0.03$ &  $  14.16\pm  0.07 $ & $ 6.14\pm  0.07$   \\  
\hline	    
\hline
\label{tab.RH}   
\end{tabular} 
\end{center}  
\end{table*}

A direct scaling between $P_R-R_H$ for radio halos is not reported in the literature.
We want to check the existence of a $P_{R}-R_H$ correlation
by making use of directly measurable quantities, such as the power and the radius at 1.4 GHz.
In the present literature it is customary to use the Largest Linear Size (LLS), 
obtained from the Largest Angular Size (LAS) measured on the radio images as the largest extension of the $2\sigma$ or $3\sigma$ contour level, as a measure of the radio emitting region (\eg Giovannini \& Feretti 2000; Kempner \& Sarazin 2001). 
Since a fraction of radio halos in our sample is characterized by a non--spherical morphology, meaning a non-circular projection on the plane of the sky, an adequate measure of a radio halo's size can be obtained by modelling the emitting volume with a spherical region of radius $R_H=\sqrt[]{R_{min}\times R_{max}}$, $R_{min}$ and $R_{max}$ being the minimum and maximum radius measured on the 3$\sigma$ radio isophotes. In this way we have 
derived the $R_H$ values for all 15 radio halos, as reported in Tab.\ref{tab.RH}, by  making use of the most recent radio maps available in literature. 
In Fig.\ref{Lr_RH} we report $P_{1.4}$ versus $R_H$ for our sample.
We find a clear trend with $R_H$ increasing with $P_{1.4}$, \ie the more extended radio halos are also the most powerful. The best-fit of this correlation is given by:

\begin{eqnarray}
\lefteqn{
\log\bigg[\frac{P_{1.4\,GHz}}{5\cdot10^{24}\,h_{70}^{-2}\,\frac{Watt}{Hz}}\bigg]=
(4.18\pm 0.68)\log\bigg[\frac{R_H}{500\,h_{70}^{-1}\,kpc}\bigg]{}}
\nonumber\\
& & {} \;\;\;\;\;\;\;\;\;\;\;\;\;\;\;\;\;\;\;\;\;\;\;\;\;\;\;\;\;-(0.26\pm 0.07)
\label{Eq.Lr_RH}
\end{eqnarray}

\noindent A Spearman test yields a correlation coefficient of $\sim 0.84$ and
a $s=0.00011$ significance, indicative of a relatively strong correlation.

\subsubsection{Uncertainties in the measure of the size of radio halos}

The dispersion of the $P_{1.4}-R_H$ correlation is relatively large, a factor of $\sim 2$ in $R_H$, and this may be due to the errors associated with the measure of $R_H$. Indeed, radio halos are low brightness diffuse radio sources which fade away gradually, until they are lost below the noise level of a given observation. Thus, the measure of a physical size is not obvious and, in any case, it needs to be explored with great care. 
However, what is important here is not so much the precise measure of $R_H$ for each radio halo, 
but rather the avoidance of selection effects which might force a correlation.

In principle the sensitivity in the different maps may play a role because the most
powerful radio halos are also the most bright ones (Feretti 2005), and thus 
they might appear more extended then the less powerful radio halos in the radio maps. 
To check if this effect is present, in Fig.\ref{brillanza} we plot the ratios between the average surface brightness of each radio halo in our sample and the rms of each map used to get $R_H$.
It is clear that there is some scattering in the distribution which would yield a corresponding dispersion in the accuracy of $R_H$, however, and most importantly, the ratios are randomly scattered, and there is no trend with $R_H$, \ie fainter radio halos are usually imaged with a higher sensitivity and thus the $P_{1.4}-R_H$ correlation cannot be forced by the maps used to derive $R_H$.

An additional effort in assessing the reliability of $R_H$ (and of the $P_{1.4}-R_H$ correlation) would be to measure the radial
brightness profile of regular radio halos which are not severely affected by powerful and extended
radio sources. In our sample it is feasible to obtain accurate radial profiles from available data for the following radio halos: A2163, A2255, A2744, A545 and A2319.
We take the data at 1.4 GHz (Feretti et al. 2001, Govoni et al. 2005, Govoni et al. 2001a, Bacchi et al. 2003, Feretti et al. 1997, respectively), and use the software package SYNAGE++ (Murgia 2001) to extract the radial brightness profiles, after subtraction of the embedded radio sources.

\begin{figure}
\centerline{\psfig{figure=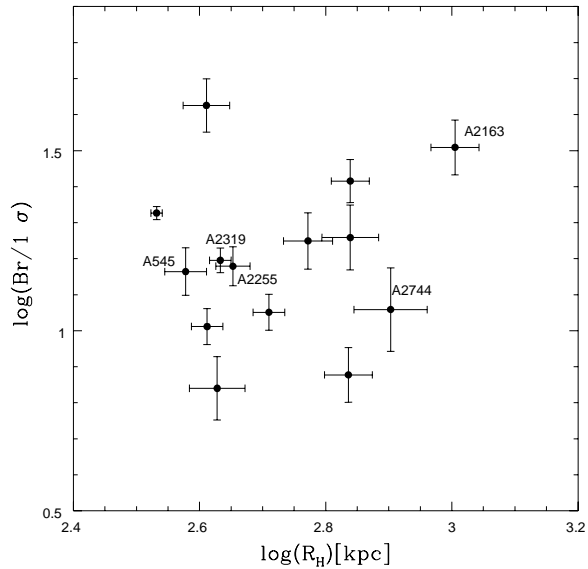,width=8cm}}
\caption[]{Ratios between the average surface brightness of each radio halo and 
the corresponding 1$\sigma$ noise level from the radio maps. The five most regular 
radio halos are earmarked.}
\label{brillanza}
\end{figure}

\begin{figure*}
\includegraphics[width=0.33\textwidth]{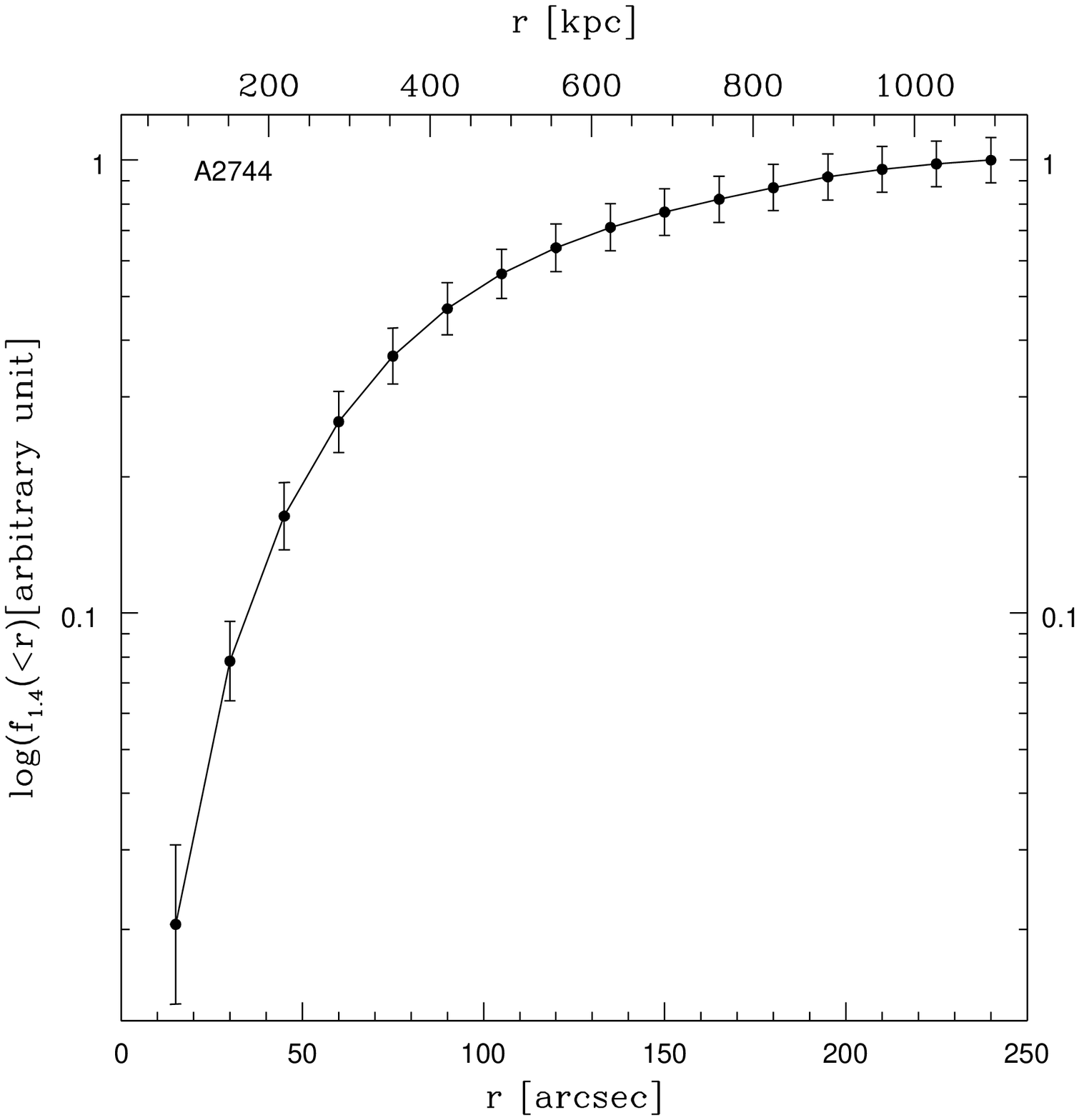}
\includegraphics[width=0.33\textwidth]{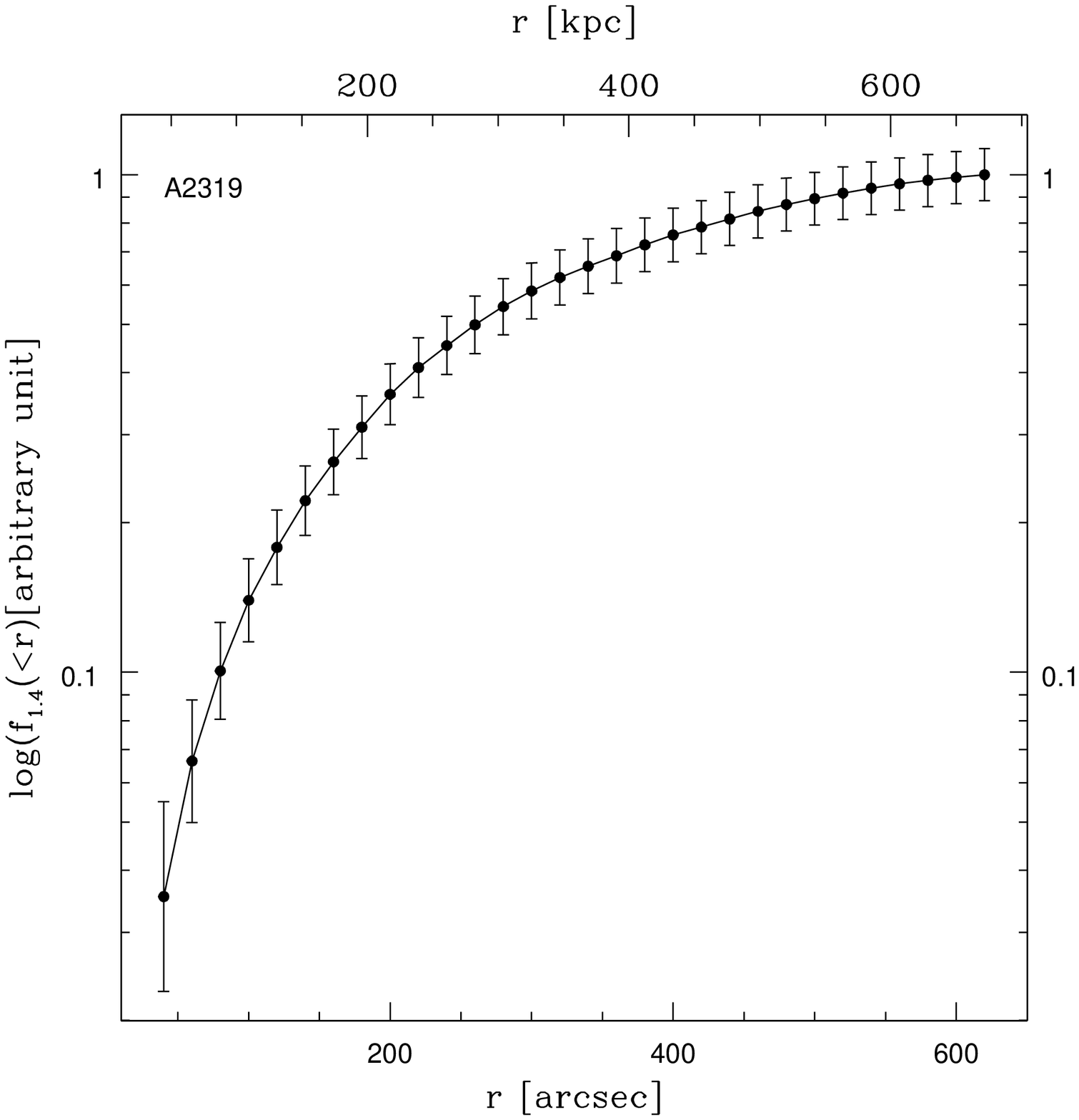}
\includegraphics[width=0.33\textwidth]{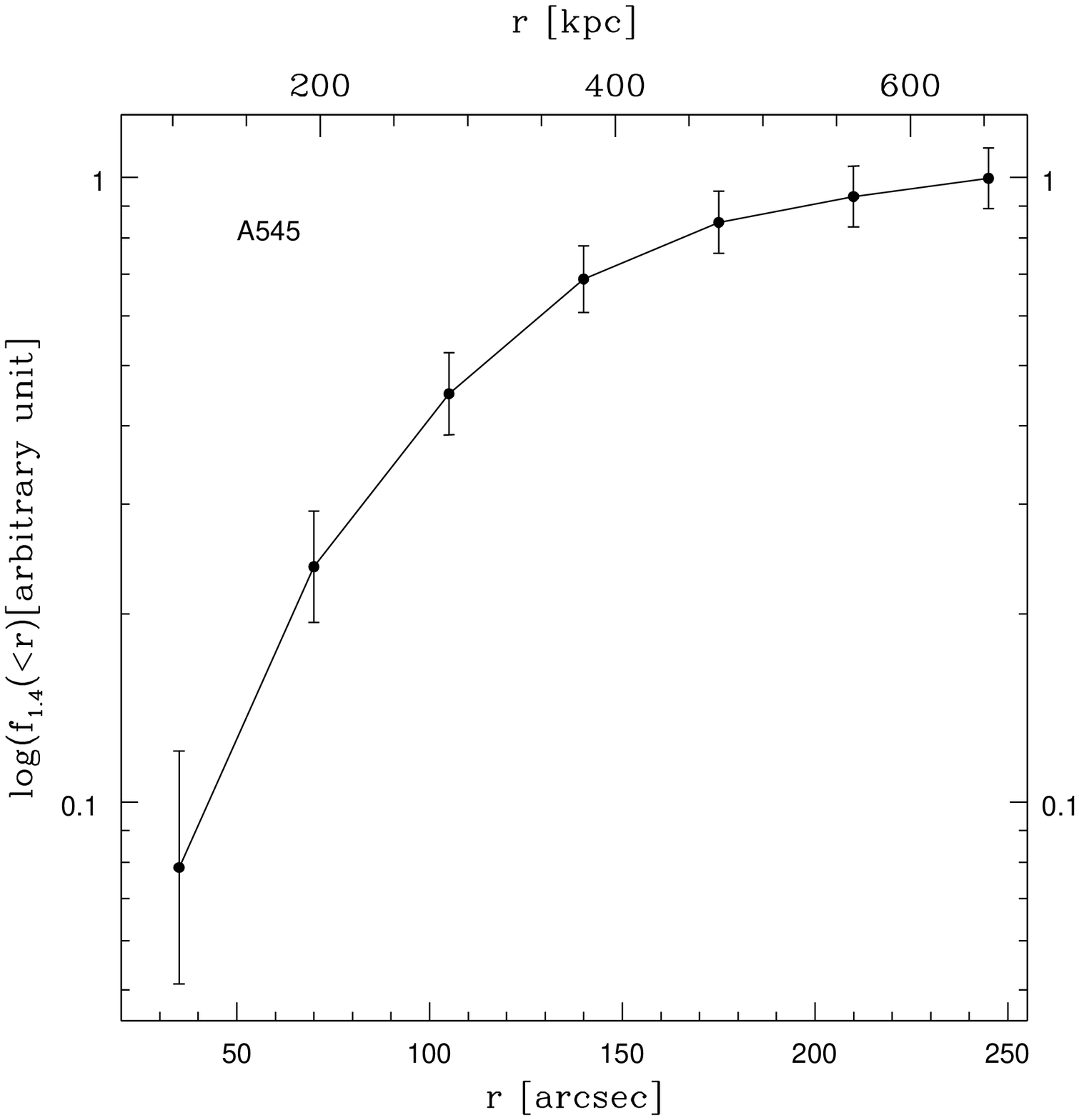}
\includegraphics[width=0.34\textwidth]{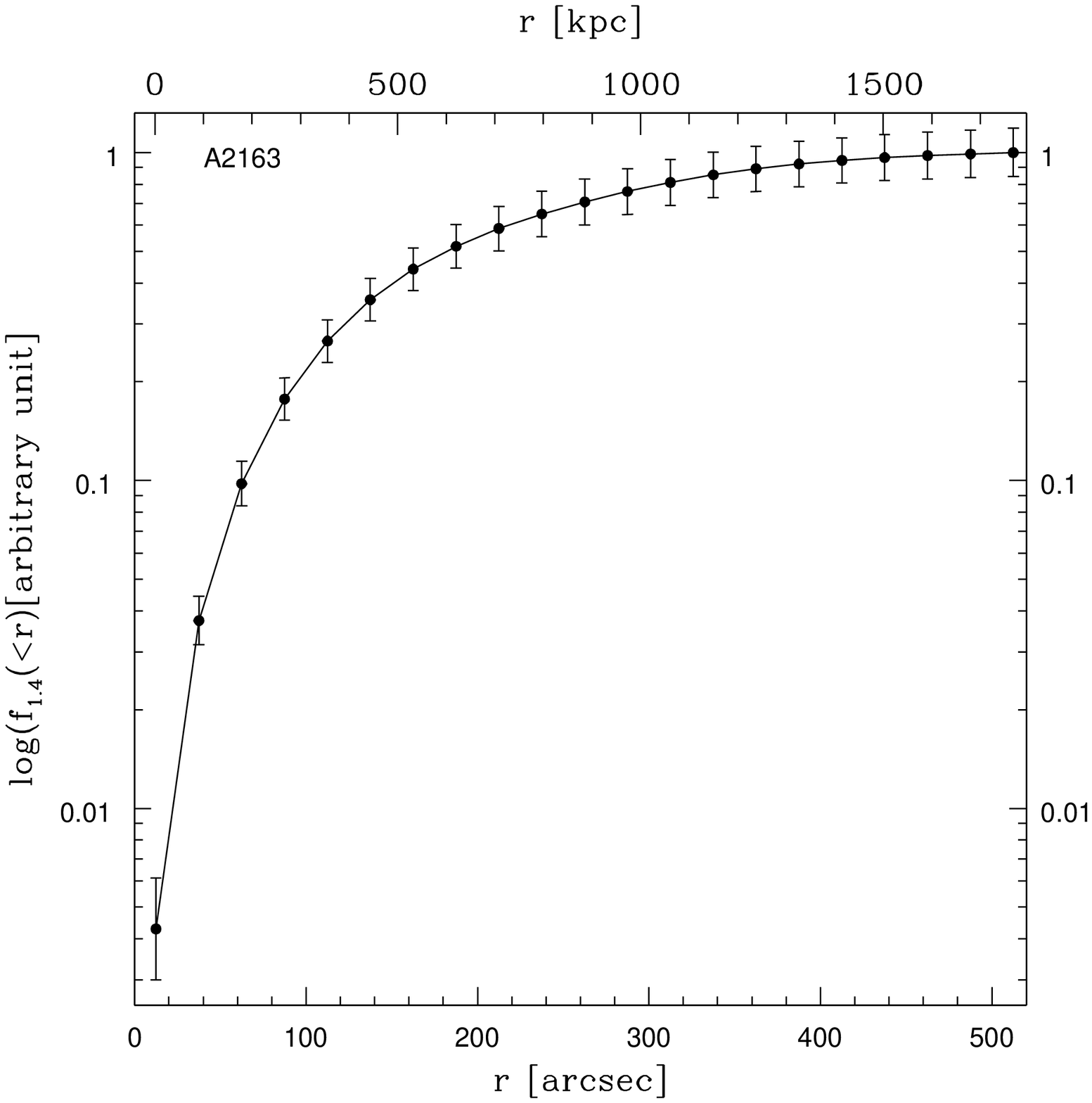}
\includegraphics[width=0.34\textwidth]{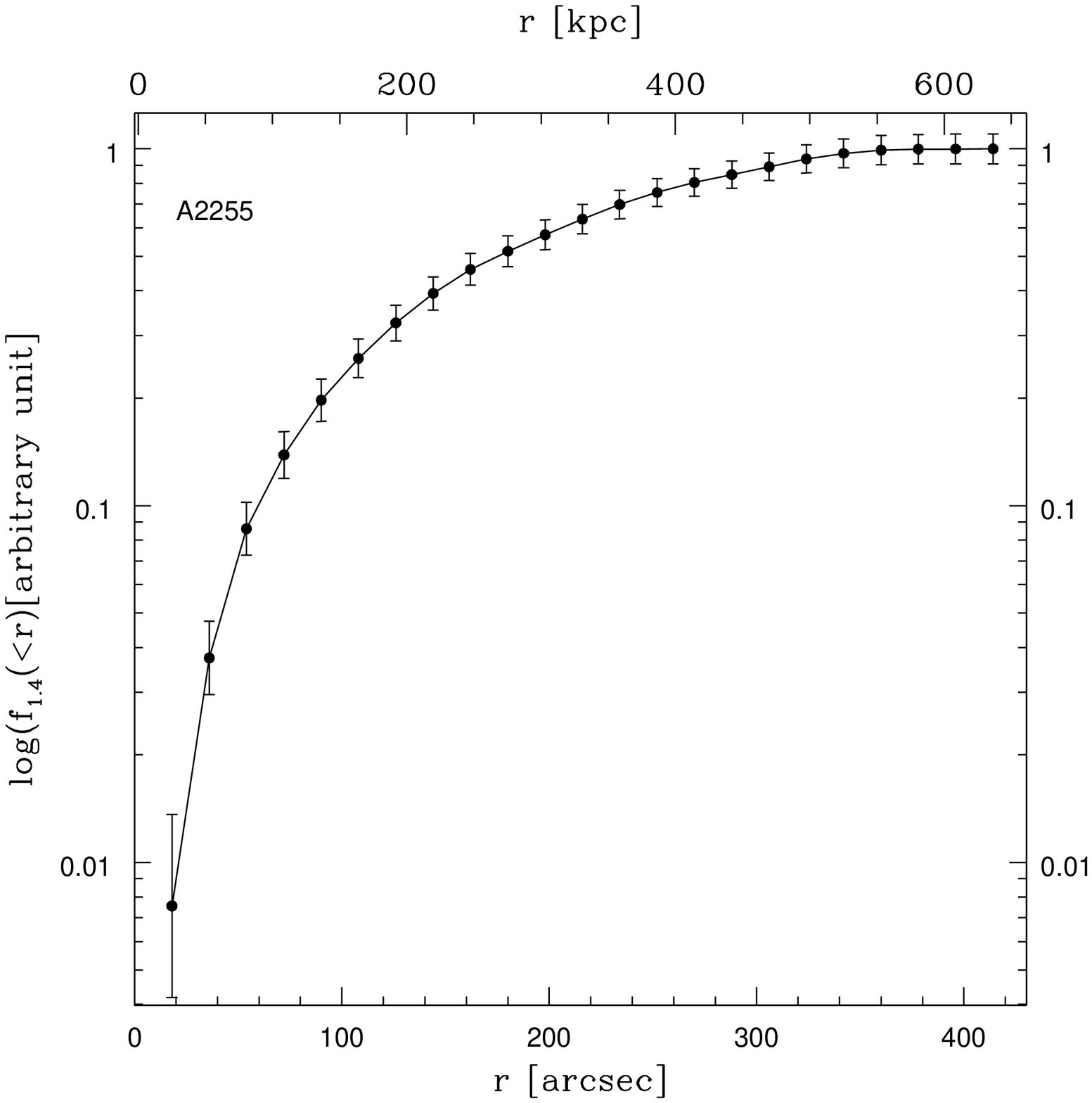}
\caption[]{Integrated radial brightness profiles of the cluster radio halos in
A2744, A2319, A545, A2163 and A2255 (from the top left to the bottom right corner).
The errors in the profiles (which are in the range 5-10\%) include the uncertainties in the sources subtraction and the statistical errors (note that in this integral presentation the errors are not independent).}
\label{profili}
\end{figure*}

In Fig.\ref{profili} we report the integrated brightness profiles of these radio halos. It is seen that the profiles flatten with distance from the respective clusters centres, indicating that basically all the extended radio 
emission is caught and that it is possible to extract an accurate physical size.  
In Fig.\ref{R_85} we report for these 5 radio halos the comparison between $R_H$, estimated directly from 3$\sigma$ radio isophotes (see the above definition), and $R_{85}$ and $R_{75}$, \ie the radii respectively containing the 85\% and 75\% of the flux of the radio halos. 
We apply the same procedure also to the case of the Coma cluster at 330 MHz for which a brightness profile and radio map were already presented in the literature (Govoni et al 2001b). For Coma at 330 MHz we find $R_H\sim\,520\,h_{70}^{-1}\,kpc$ and $R_{85}\sim\,610\,h_{70}^{-1}\,kpc$, which set Coma in a configuration similar to that
of the other clusters in Fig.\ref{R_85}.

The linear, almost one-to-one correlation between $R_H$ and $R_{85}$ and the relatively small dispersion, consistent with the uncertainties in the profiles due to source subtraction, 
prove that our definition of $R_H$ is a simple but representative estimate of the physical size of radio halos.

\begin{figure}
\centerline{\psfig{figure=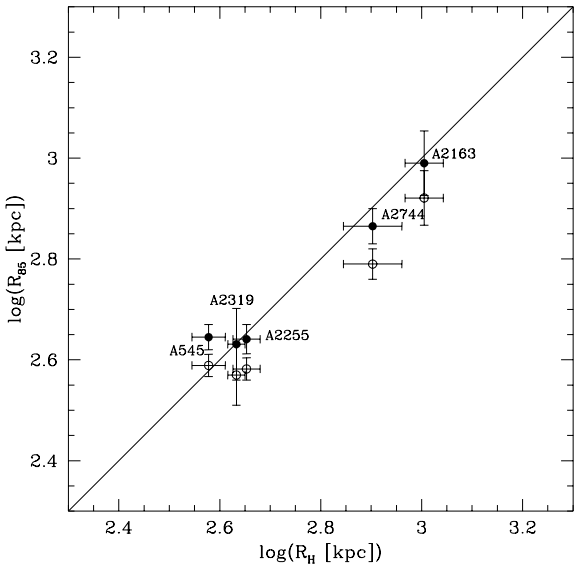,width=8cm}}
\caption[]{Radius enclosing the 85\% (filled circles) and the 75\% (open circle)
of the total radio flux at 1.4 GHz obtained by the profiles (Fig.\ref{profili}) 
versus $R_H$ estimated directly from the radio 
maps at 1.4 GHz.}
\label{R_85}
\end{figure}

We note that the sensitivities of the radio maps, the physical sizes $R_{85}$ and powers $P_{1.4}$ of the 5 regular halos are representatives of the values encompassed by the full radio halo sample. Moreover, for these 5 radio halos alone we find $P_{1.4}\propto R_{85}^{4.25\pm0.63}$, fully consistent with the $P_{1.4}-R_H$ correlation obtained for the total sample.

\subsubsection{Possible biases in the selection of the sample}

Since the $P_{1.4}-R_H$ correlation is the driving correlation, one has to 
check whether this correlation may not be forced by observational biases 
due to the selection of the radio halo population itself. 
Indeed the great majority of these radio halos have been discovered 
by follow-ups of candidates, mostly identified from the NVSS
which is surface brightness-limited for resolved sources\footnote{The rms brightness fluctuations in the NVSS are $0.45$ mJy/beam (beam=45$\times$45 arcsec, Condon et al. 1998); the NVSS is sensitive to radio sources with size $< 10'-15'$ (appropriate for radio halos at $z>0.05$; GTF99).} and this may introduce biases in the selected sample.

The upper bound of the correlation is likely to be solid: 
objects as powerful as those at the upper end of the correlation 
($\log P_{1.4}\geq 25$) but with small $R_H$ (similar to that of 
radio halos in the lower end of the correlation) should appear in the NVSS up to the largest 
redshifts of the sample, since, even at $z \sim 0.3$, they should be
$\geq$10 times brighter than the low power radio halos in the correlation 
and extended ($\sim 2.5'$). As a matter of fact A545 (z=0.15) and A520 
(z=0.2), which are among the smaller radio halos in our sample,  
are already detected in the NVSS up to a redshift 0.2 and there is no reason why
objects with similar extension, but $\sim 8-10$ times brighter than A545 and A520, 
should not have been detected at $z\leq0.3$.

The lower bound of the correlation deserves much care since the brightness limit
of the NVSS may play some role. 
It is clear that present surveys may significantly affect the selection of the
faint end of the radio halo population. However, Feretti (2005) and Clarke (2005), have already concluded that the typical brightness of the powerful and giant radio halos are well above the detection limit.

\begin{figure}
\centerline{\psfig{figure=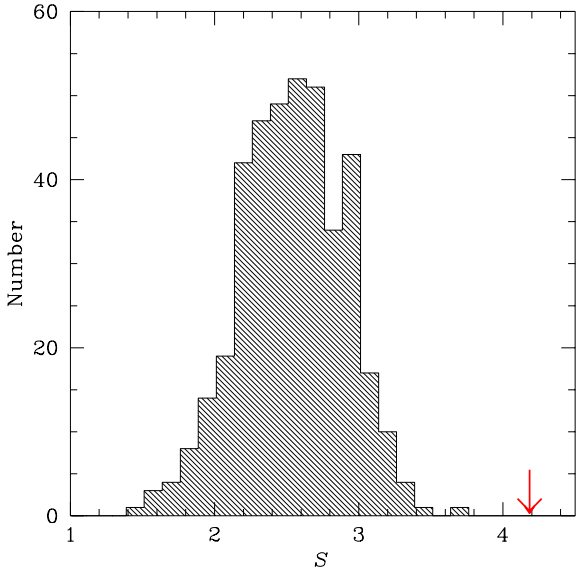,width=8cm}}
\caption[]{Distribution of the slopes, {\it S}, of the $P_{1.4}-R_H$ correlation
obtained with our Monte Carlo procedure (with 400 trails). 
The arrow indicate the value of the observed best-fit slope $\simeq 4.18$.}
\label{montecarlo}
\end{figure}

In any case, a brightness limit should drive a $P_{1.4}\propto R_H^2$ correlation,
much flatter then the observed one. In order to provide  a further compelling 
argument against observational biases, we have run Monte Carlo simulations. 
To this end we have randomly extracted brightness values of hypothetical radio halos
within a factor of $\sim 5$ interval (consistently with the range spanned in our sample) 
above a given minimum brightness and each time randomly assigned $R_H$ and z 
among the observed values.
In Fig.\ref{montecarlo} we report the distribution of the $P_{1.4}-R_H$ slopes obtained with
our Monte Carlo procedure and note that this distribution is peaked around 
$\sim 2.5$ with a dispersion of $\pm 0.4$ (this is somewhat steeper than the expected $P_{1.4}\propto R_H^2$ due to the well known redshift effect, however small given the small redshift range of our sample). The values of the slopes from the Monte Carlo procedure are far from the observed value (Fig.\ref{montecarlo}) and a statistical test allows us
to conclude that the probability that the observed $P_{1.4}-R_H$ correlation is forced by observational biases is $\ltsim0.05\%$.

\subsection{Geometrical $M_H-R_H$ scaling for radio halos}

The existence of a possible tight scaling between the size of radio halos and the cluster
mass within the emitting region is not reported in the literature. Yet an observational
$M_H-R_H$ scaling may be important to relate virial quantities $\sigma_v^2=G\,M_v/R_v$
($\approx G\,M_H/R_H=\sigma_H^2$) with quantities ($R_H$ and $M_H$) which refer to the emitting region, and to test simple model expectations (Sect.~2.2).

At this stage of the paper, the main difficulty concerns the 
measure of the cluster mass inside a volume of size $R_H$.
Here the only possibility is to use the X-ray mass determination based 
on the assumption of hydrostatic equilibrium. 
Nevertheless, radio halo clusters are not well relaxed systems and thus the assumption of 
hydrostatic equilibrium and spherical symmetry may introduce sizeable errors in the mass determination. 
Several numerical simulation studies, which have been undertaken in order 
to determine whether the above assumptions introduce significant 
uncertainties in the mass estimates, indicate that in the case of merging 
clusters the hydrostatic equilibrium method might lead to errors  
up to 40\% of the true mass, which can be either overestimated
or underestimated (\eg Evrard et al. 1996; R\"ottiger et al. 1996; 
Schindler 1996; Rasia et al. 2006). This would cause an unavoidable
scattering in the determination of the mass in our sample, although
there are indications that a better agreement between the gravitational lensing, 
X-ray and optically determined cluster masses is achieved on scales larger than 
the X-ray core radii (\eg Wu 1994; Allen 1998; Wu et al. 1998), 
which is the case under consideration ($R_H>r_c$).

However, what is important here is that the mass determination does not introduce systematic errors which depend on the mass itself and which may thus affect the {\it real} trend of the $P_{1.4}-M_H$ correlation.
We thus compute the total gravitational cluster mass within the radius $R_H$ as:

\begin{equation}
M_{H}=M_{tot}(<R_H)=\frac{3K_{B}T R_H^{3}\beta}{\mu m_p G}\left(\frac{1}{R_H^2+r_c^2}\right)
\label{MH}
\end{equation}

\begin{figure}
\centerline{\psfig{figure=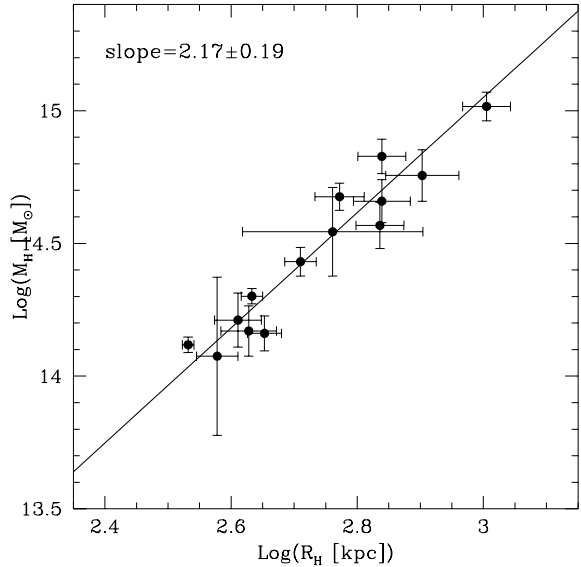,width=8cm}}
\caption[]{$M_H$ versus $R_H$ for giant radio halos. 
The best-fit power-law and the value of the slope are
also reported in the panel.}
\label{observed1}
\end{figure}

\begin{figure}
\centerline{\psfig{figure=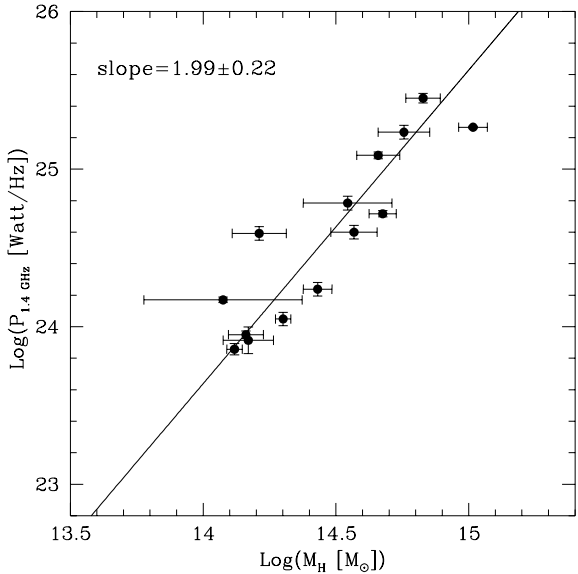,width=8cm}}
\caption[]{$P_{1.4}$ versus $M_H$ for giant radio halos. The best-fit power-law and its slope
are also reported in the panel.}
\label{observed2}
\end{figure}

\begin{figure}
\centerline{\psfig{figure=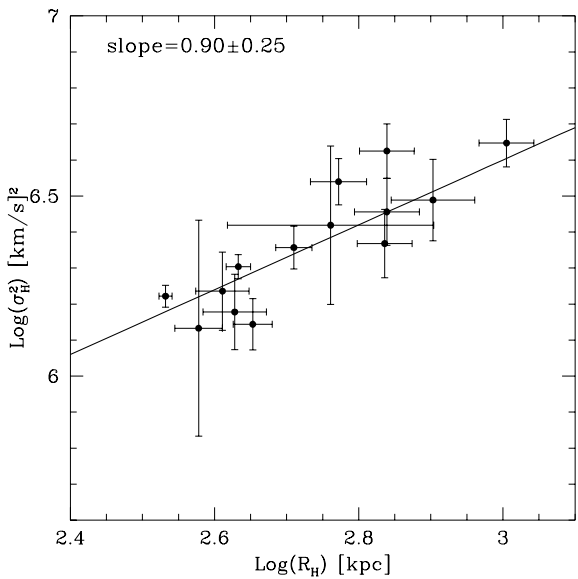,width=8cm}}
\caption[]{Square of the velocity dispersion inside $R_H$ versus $R_H$. 
The best-fit power-law and its slope
are also reported in the panel.}
\label{observed4}
\end{figure}

\noindent where $r_c$ is the core radius, $T$ the isothermal gas temperature and $\beta$ the ratio between the kinetic energy of the dark matter and that of the gas ($\beta$-model; \eg Sarazin 1986). 
We have excluded from our analysis A2254 for which no information on the $\beta$-model is available. For the remaining 14 clusters references are given in CBS06. From Eq.\ref{MH} one has that $M_H\propto R_H$ for $R_H>>r_c$ and 
$M_H\propto R_H^3$ for $R_H<<r_c$. In Fig.\ref{observed1} we plot 
$R_H$ versus $M_H$ for our sample: we find $M_H\propto R_H^{2.17\pm0.19}$,
which falls in between the above asymptotic expectations.

\subsection{Radio power versus mass and velocity dispersion}

In principle, the two correlations discussed so far for giant radio halos, $P_{1.4}-R_H$ and $M_H-R_H$, imply 
the existence of correlations between $P_{1.4}-M_H$ and $P_{1.4}-\sigma_H$. In particular $P_{1.4}$ is expected to roughly scale as $M_H^{1.9-2}$. In Fig.\ref{observed2} we report $P_{1.4}$ versus $M_H$ for our sample together with the best-fit: $P_{1.4}\propto M_H^{1.99\pm0.22}$, which is indeed in line with the above expectation. A Spearman test of this correlation yields a correlation coefficient of $\sim 0.91$ and $s=7.3\cdot 10^{-6}$ significance, indicative of a very strong correlation.

$P_{1.4}$ is expected to scale with $\sigma_H$ and we found for our sample a best-fit correlation: $P_{1.4}\propto (\sigma_H^2)^{4.64\pm 1.07}$; a Spearman test yields a correlation coefficient of $\sim 0.89$ and to $s=2\cdot 10^{-5}$ significance, indicative of a very strong correlation. 

Finally, as a by-product of all the derived scalings, it is worth noticing that also a trend between $R_H-\sigma_H$ is expected (Fig.\ref{observed4}). This finding might also be tested by observations in the optical domain which can directly constrain the velocity dispersion.

\begin{figure}
\centerline{\psfig{figure=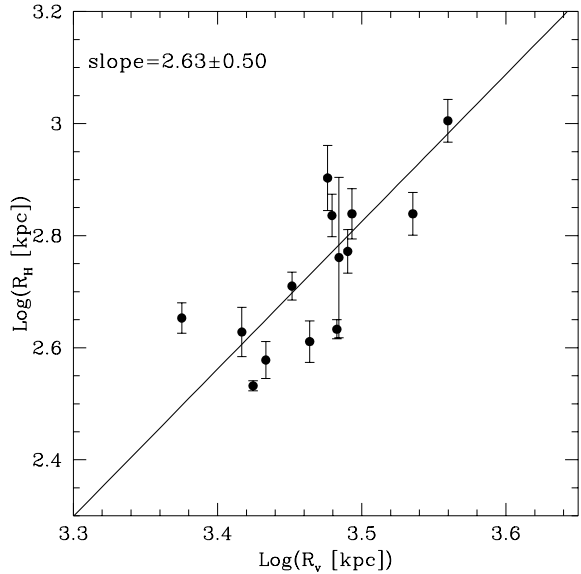,width=8cm}}
\caption{$R_H$ versus virial radius,$R_v$, of hosting clusters estimated from the $Lx-M_v$ correlation (see CBS06). In the panel is also reported the best-fit correlation.}
\label{RH_Rv}
\end{figure}

\section{Implications of the derived scalings}

Given that the larger radio halos are also the most powerful ones and are hosted in the most massive clusters, we expect that the size of a giant radio halo should scale with the size of the hosting cluster. We estimate for each cluster of our sample the virial radius ($R_v$) by combining the virial mass--X-ray correlation ($M_v-L_X$; CBS06) and the virial radius-virial mass relation (e.g. Kitayama \& Suto 1996). This method allows to reduce the effect of scattering due to the uncertainties in the mass measurements (and thus in the $R_v$) of merging galaxy clusters (see discussion in CBS06). In Fig.\ref{RH_Rv} we plot $R_H$ versus $R_v$ for our sample. The best fit gives $R_H\propto R_v^{2.63\pm 0.50}$ , i.e. {\em a pronounced non-linear increase of the size of the radio emitting region with the virial radius}. A Spearman test yields a correlation coefficient of $\sim 0.74$ and $s=0.0023$ significance, indicative of a relatively strong correlation, albeit less strong than the {\it others correlations found in this paper}. 
 
Given that massive clusters are almost self similar (e.g. Rosati et al. 2002) one might have expected that $R_H$ scales with $R_v$ and that the radial profiles of the radio emission are self-similar. On the contrary, our results prove that self-similarity is broken in the case of the non-thermal cluster components. This property of radio halos was also noticed by Kempner \& Sarazin (2001), which used a sample of radio halos taken from Feretti (2000) and found evidence for a trend of the Largest Linear Size, LLS, with the X-ray luminosity in the 0.1-2.4 keV band, $LLS\propto L_x^{1/2}$, while a flatter scaling, $LLS\propto R_v \propto L_X^{1/6}$ is expected in the case of a self-similarity. Their results imply $R_H\propto R_v^{3}$; if one takes $R_H\approx LLS$, this is substantially in line with our findings. It is also worth noticing that X-ray--radio comparison studies of a few radio halos indicates that the profile of the radio emission is typically broader than that of the thermal emission (\eg Govoni et al. 2001b).
The two ingredients which should be responsible for the break of the self--similarity are the distributions of relativistic electrons and magnetic fields. In MHD cosmological simulations (Dolag et al. 2002, 2005) it is found that the magnetic field strength in cluster cores increases non-linearly with cluster mass (temperature). This implies that the radio emitting volume should increase with cluster mass because the magnetic field at a given distance from the centre increases with increasing the mass. A detailed analysis of the magnetic field profiles of massive clusters from MHD simulations could be of help  in testing if the magnetic field is the principal cause of the break of the self-similarity.

\section{Particle re-acceleration model and observed scalings}

Although we have been guided by the analysis of Eq.\ref{epsilonsynint} to predict the existence of scaling relationships, the observed correlations derived in Sec. 3 are actually independent from the form of this equation. 
To test Eq.\ref{epsilonsynint} against the observed quantities of our sample of radio halos we make use of the monochromatic $P_{1.4}$ instead of the unavailable bolometric $P_{R}$. This is possible because the typical spectral shape of radio halos is $\alpha_r\approx 1.1-1.2$ ($P(\nu)\propto \nu^{-\alpha_r}$) and thus the K-correction is not important (CBS06).

\begin{figure}
\centerline{\psfig{figure=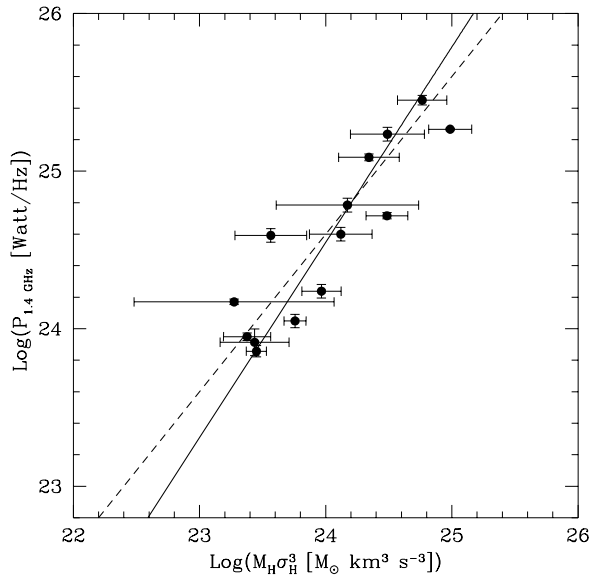,width=8cm}}
\caption[]{$P_{1.4}$ versus $M_H\,\sigma_H^3$. The best-fits correlations (solid line) and the predicted scaling with $\mathcal{F}\sim$ constant (dashed line) are reported.}
\label{P_MHsigma3}
\end{figure}

In Fig.\ref{P_MHsigma3} we report $P_{1.4}$ versus $M_H\,\sigma_H^3$. The best fit gives $P_{1.4}\propto (M_H\,\sigma_H^3)^{1.24\pm 0.19}$. The observed scaling is slightly steeper,
but still in line with the linear scaling expected from Eq.\ref{epsilonsynint} for $\mathcal{F}$ constant (dashed line). As already discussed in Sec.~2.2 $\mathcal{F}$ is constant for $B_H^2>>B_{cmb}^2$ or in the case in which the rms magnetic field in the radio halo region is quite independent from the cluster mass (small $b_H$), while formally a non--constant $\mathcal{F}$ always implies a steepening of the $P_{1.4}-M_H\,\sigma_H^3$ scaling. Namely, in the case of 
$\sim \mu$G magnetic fields, by combining Eq.\ref{epsilonsynint} with the observed $M_H-R_H$ correlation (Sec.~3.2, Fig.\ref{observed1}), one has that the best-fit in Fig.\ref{P_MHsigma3} is fulfilled by the model expectations for $0.05\leqslant b_H\leqslant 0.39$.

In principle the fit can be used to set constraints on the values of the theoretical parameters entering the normalization of Eq.\ref{epsilonsynint}, (namely $\epsilon_{CR}/\epsilon_{th}$, and the fraction of the PdV work which goes into turbulence), but we will not pursue this any further here (see CB05 for a discussion).

It is important to stress that not only the trend in Fig.~\ref{P_MHsigma3}, but also the existence of the correlations found in Sec.~3 could have been predicted on the basis of the re-acceleration model (Sec.~2, Eqs.~\ref{eq.RH},~\ref{eq.MH},~\ref{eq.sigmaH}) under the very reasonable assumption that $M_H\propto R_H^{\alpha}$. Indeed, if one uses the observed scaling $M_H \propto R_H^{2.17\pm 0.19}$ to fix the parameter $\alpha$, from Eq.\ref{epsilonsynint}, and assuming the most simple case in which $\mathcal{F}$ is constant, one finds $P_{1.4}\propto R_H^{3.9}$ and $P_{1.4}\propto M_H^{1.8}$, which are actually consistent (within the dispersion) with the observed correlations (Sec.~3); as in the case of the trend in Fig.~\ref{P_MHsigma3}, an even better fulfillment of all these correlations is obtained for a slightly non-constant $\mathcal{F}$.

A relevant point which derive from the comparison of model expectation and observed correlations (unless $B_H^2>>B_{cmb}^2$) is that, at least under our simplified approach (Sec.~2.2), $B_H$ does not critically depend on cluster mass inside $R_H$ and that radio halos might essentially select the regions of the cluster volume in which the magnetic field strength is above some minimum value (say $\sim\,\mu$G level). It is important to note that a roughly constant $B_H$ with cluster mass does not contradict the scaling of B, averaged in a fixed volume, with cluster mass (or temperature) found in the MHD simulations (B within the cluster core radius, $r_c\sim 300\,h_{70}^{-1}\,kpc$), and also found in CBS06 (B averaged within a fixed region of $\sim\,720\,h_{70}^{-1}\,kpc$ size), because the magnetic field $B_H$ is averaged over a volume of radius $R_H$ that becomes substantially larger than the core radius with increasing the cluster mass ($R_H/r_c$ goes from $\sim 1.1$ to
$\sim 3$ with increasing cluster mass in our sample).

\section{Summary \& Conclusions}
The particle re-acceleration model is a promising possibility to explain the origin and properties of the giant radio halos (\eg Brunetti 2004; Blasi 2004; Hwang 2004; Feretti 2005, for recent reviews).\\
\noindent $\bullet$ In its simplest form, as assumed here (Sect.2), it predicts 
a very simple relationship (Eq.\ref{epsilonsynint}) between the total radio power $P_R$, the total mass $M_H$ within the radio halo, the gas velocity dispersion $\sigma_H$ and the average magnetic field $B_H$.   
Under the assumption of a tight scaling between $M_H$ and the size $R_H$, and that the gas is in gravitational equilibrium, Eq.\ref{epsilonsynint} naturally translates into simple scaling relations: $P_R-R_H$, $P_R-M_H$, and $P_R-\sigma_H$ (Eqs.~\ref{eq.RH},~\ref{eq.MH},~\ref{eq.sigmaH}).\\

\noindent Motivated by the above theoretical considerations, we have searched for the existence of this type of correlations by analyzing a sample of 15 galaxy clusters 
with giant radio halos. 
A most important point here is the measure of the size $R_H$, in itself a
non-trivial matter, since the brightest radio halos may appear more extended in the radio 
maps and this might force artificial correlations with radio power. A careful analysis
of published 15 GHz radio maps of the radio halos of our sample shows that this effect is not present (Sec.3.1.1). 
From the same data set we derive a meaningful estimate of the radius for each radio halos. We also show that our procedure leads to estimates fully consistent with the measurements from the brightness profiles worked out from the data for the five most regular radio halos; this consistency holds over the total range spanned by $R_H$ in our sample (Sect. 3.1.1).

\noindent $\bullet$ We obtain a good, new correlation (correlation coefficient $\sim 0.84$) 
between the observed radio power at 1.4 GHz and the measured size of the 
radio halos in the form $P_{1.4}\propto R_{H}^{4.18\pm 0.68}$ (Sect.3.1). 
In Sect.3.1.2 we discuss in detail several selection effects which might affect this correlation and conclude that it is unlikely that the observed correlation is 
driven by observational biases.

$\bullet$ We address observationally also the presence of a tight scaling between $M_H$ and $R_H$ and this allows us to relate virial quantities to quantities in the emitting region.

\noindent $\bullet$ The presence of the $P_R-R_H$ and $M_H-R_H$ correlations implies also other correlations. We derive relatively strong correlations  (Sect. 3.3) in the form: $P_{1.4}\propto M_{H}^{1.99\pm 0.22}$ and $P_{1.4}\propto (\sigma_{H}^2)^{4.64\pm 1.07}$, and, as a byproduct, 
also $\sigma_H^2\propto R_H^{0.90\pm 0.25}$.\\

\noindent A correlation between the size $R_H$ and the cluster virial radius, $R_v$, is qualitatively expected in the framework of the particle re-acceleration model.  

\noindent $\bullet$ In Sec.~4 we compare $R_H$ vs. $R_v$ for our sample of clusters with giant radio halos, obtaining the {\it non-linear} trend $R_H\propto R_v^{2.63\pm 0.50}$, \ie the fraction of the cluster volume that is radio emitting significantly increases with the cluster mass. This break of the self-similarity, in line with previous suggestions (\eg Kempner \& Sarazin 2001), points to the changing distributions of the magnetic fields and relativistic electrons with cluster mass and, as such, is  potentially important in constraining the physical parameters entering the hierarchical formation scenario, such as the turbulence injection scale and the magnetic field strength and profile. Finally, we note that, by combining the $R_H-R_v$ and $P_{1.4}-R_H$ correlations, one easily derives $P_{1.4}\propto M_v^{3}$, which is consistent with previous findings ($P_{1.4}\propto M_v^{2.9\pm 0.4}$; CBS06).\\

\noindent $\bullet$ These observed correlations are well understood in the framework of the particle re-acceleration model. Indeed, we show that the theoretical expectation (Eq.\ref{epsilonsynint}) is consistent with the data (see Fig.\ref{P_MHsigma3}).
Assuming a simple constant form for $\mathcal{F}$ in Eq.\ref{epsilonsynint} and 
the observed $M_H-R_H$ scaling, which is necessary to fix the model parameter $\alpha$ (Sect.2), the model expectations (Eqs.~\ref{eq.MH},~\ref{eq.RH},~\ref{eq.sigmaH}) naturally translates into $P_{1.4}\propto R_H^{3.9}$, $P_{1.4}\propto M_H^{1.8}$ and $P_{1.4}\propto (\sigma_H^{2})^{3.4}$ correlations, all consistent (within the dispersion) with the observed correlations; an even better fulfillment of all these correlations is obtained for a slightly non-constant $\mathcal{F}$, which corresponds to $\approx\mu G$ field in the radio halo region.
Unless it is $B_H^2>>B_{cmb}^2$, from the comparison of model expectations and observations we conclude that $B_H$ should not strongly depend on $M_H$, and thus in our simplified scenario (Sec.~2.2) radio halos essentially trace the regions of $\approx\,\mu$G fields in galaxy clusters in which particle
acceleration is powered by turbulence.\\

\noindent To conclude, the particle re-acceleration model, closely linked to the development of the turbulence in the hierarchical formation scenario, appears to provide a viable and basic physical interpretation for all the correlations obtained so far with the available data for giant radio halos. Future deep radio surveys and upcoming data from LOFAR and LWA will be crucial to improve the statistics and to provide further constraints on the origin of radio halos.

\section*{Acknowledgments}
RC acknowledge the MPA in Garching for the hospitality during
the preparation of this paper. We thank Matteo Murgia for 
the use of the SYNAGE++ program. We thank Luigina Feretti for providing the data for
A2163, A545 and A2319, and Marco Bondi for useful discussions.
This work is partially supported by MIUR and INAF under grants PRIN2004, PRIN2005 and
PRIN-INAF2005.

\clearpage

\end{document}